\documentclass[12pt]{iopart}
\usepackage{epsf, graphicx,iopams} 
\usepackage[squaren]{SIunits}

\def\lsim{\mathrel{\rlap{\lower4pt\hbox{\hskip1pt$\sim$}}
    \raise1pt\hbox{$<$}}}

\begin{document}

\title[GPU-accelerated inspiral searches]{Application of Graphics Processing Units to Search Pipeline for Gravitational Waves from Coalescing Binaries of Compact Objects}

\author{Shin Kee Chung$^{1}$, Linqing Wen\footnote{Also, International Centre for Radio Astronomy Research, UWA, 35 Stirling Hwy, Crawley, WA 6009, Australia. Email: lwen@cyllene.uwa.edu.au}$^{1}$, David Blair$^{1}$, Kipp Cannon$^{2}$, Amitava Datta$^{3}$}
\address{$^1$ School of Physics, The University of Western Australia, 35 Stirling Highway, Crawley WA 6009, Australia}
\address{$^2$ Division of Physics, Mathematics and Astronomy, California Institute of Technology, 1200 California Blvd., Pasadena CA 91125, USA}

\address{$^3$ School of Computer Science and Software Engineering, The University of Western Australia, 35 Stirling Highway, Crawley WA 6009, Australia}

\begin{center}
\begin{abstract}
We report a novel application of a graphics processing unit (GPU) for the purpose of accelerating the search pipelines 
for gravitational waves from coalescing binaries of compact objects. 
A speed-up of 16 fold in total has been achieved with an NVIDIA GeForce 8800 Ultra GPU card compared with one core of a 2.5 GHz Intel Q9300 central processing unit (CPU). 
We show that substantial improvements are possible and discuss the 
reduction in CPU count required for the detection of inspiral sources afforded by the use of GPUs.
\end{abstract}
\end{center}



\maketitle

\section{Introduction}
\label{intro}
It is an exciting time for gravitational-wave astronomy.  Several ground-based 
gravitational-wave (GW) detectors have reached (or approached) their design sensitivity, 
and are coordinating to operate as a global array. These include the three LIGO detectors 
in Louisiana and Washington states of USA, and the Virgo detector in Italy. 
The LIGO detectors have already completed their ground-breaking fifth science run.
An integrated full year's worth of science data summing up to more than 10 terabytes has been accumulated from all three interferometers in coincidence. 
Enhanced LIGO started to operate in June 2009 and plan to continue until the end of 2011 with similar sensitivity~\cite{pathtoenhanced}. 
Starting in 2011, an upgrade of Enhanced LIGO to Advanced LIGO (expected to operate in 2014)  will enable a 10-fold improvement in sensitivity, 
allowing the detectors to monitor a volume of the universe 1000 times larger than current detectors. 
The detection rate of signals from coalescing binaries of compact objects for these advanced detectors is estimated to be tens to hundreds of events per year~\cite{pathtoenhanced}. 
The detection of the first GW is virtually assured with Advanced LIGO. 

\subsection{Search for Gravitational Wave from Coalescing Compact Binaries}
Coalescing binaries of compact objects consisting of neutron stars and black holes are among the most important GW sources targeted by current large scale GW detectors~\cite{ligoanalysis, coalescingbinaries}
as these sources produce a very distinct pattern of gravitational wave. 
The optimal way to detect known waveforms in noisy data is to perform a matched filtering.
The matched filtering technique is performed by calculating the 
correlation between the gravitational wave data and a set of known 
or predicted waveform templates~\cite{matchedfiltering, finn}. 
The post-Newtonian expansion method is used to approximate 
the non-linear equations that describe the motion of coalescing binaries and 
wave generation in the creation of waveform templates~\cite{blanchet96}. 
For spinless, circular, binary systems each waveform is specified by a set of
parameters including a pair of individual masses $I = (m_1, m_2)$, constant 
orbital phase offset at formal coalescence $\alpha$ and 
effective distance $D_{\rm eff}$ from the detector.
In our tests, second order post-Newtonian orbital phases and Newtonian amplitude were used.

The application of the matched filtering technique to on-going searches for 
gravitational waves from coalescing binaries of compact objects is described 
in~\cite{ligoanalysis} and is summarized as follows.  The
template waveforms corresponding to $\alpha = 0$ and $\alpha = \pi/2$ 
form an orthonormal set~\cite{prd49}.
For a given mass pair $I = (m_1, m_2)$, the waveforms with $\alpha = 0$ and 
$\alpha = \pi/2$, denoted  $h_c^I$ and $h_s^I$ are approximately related by 
$\tilde{h}_c^I(f) = -i\tilde{h}_s^I(f)$ where $\tilde{h}$ is the Fourier transform of $h$.
Exploiting this, the matched-filter output $z(t)$ is a complex time
series defined as 
\begin{equation}
\label{eqn1}
z(t) = x(t) + \mathrm{i}y(t) = 4 \int_{0}^{\infty} \frac{{\tilde{h}_c^{I}(f)}
\tilde{s}^*(f)} {S_n(f)} \mathrm{e}^{2\pi \mathrm{i} f t} \mathrm{d}f
\end{equation}
where $S_n(f)$ is the one-sided strain noise power spectral density, $\tilde{s}^*(f)$ is
the Fourier transform complex conjugate of the detector's calibrated strain data $s(t)$,
$x(t)$ is the matched-filter response of $h^I_c$ and $y(t)$ is 
the matched-filter response of $h^I_s$. 
For initial LIGO detectors, $S_n(f)$ is defined as
\begin{equation}
\label{sfequation}
S_n(f) = \left(\frac{4.49f} {f_0}\right)^{-56} + 0.16\left( \frac{f} {f_0} \right)^{-4.52} + 0.52 + 0.32\left( \frac{f} {f_0} \right)^2
\end{equation}
where $f_0 = 150Hz$~\cite{sfformula}. In practice, the noise power spectral density from the
Science Requirement Document is used~\cite{ligosrd}.
Maximizing over the coalescence phase $\alpha$, the signal-to-noise ratio (SNR) 
$\rho(t)$ is the absolute value of the scalar
product between a normalized template and the detector output in 
frequency domain~\cite{ligoanalysis}
\begin{equation}
\label{eqn2}
\rho(t) = \frac{|z(t)|} {\sigma}
\end{equation}
where the normalization factor $\sigma$ is calculated from the variance
\begin{equation}
\label{eqn3}
\sigma^2 = 4 \int_{0}^{\infty} \frac{|\tilde{h}_c^{I}(f)|^2} {S_n(f)}
\mathrm{d}f.
\end{equation}
For stationary and Gaussian noise, this $\rho$ is the optimal detection statisticc 
for a single detector.

The number of templates needed depends on the parameter volume needed to be searched. 
The two masses of the compact binary objects were used as the main parameters 
in our experiment as we were focusing on spinless and circular binaries of compact objects. 
The low frequency cutoff was set to be 40 Hz while high frequency cutoff is the 
Nyquist frequency (half of the sampling frequency, 2 kHz in our case) or
the frequency of innermost stable circular orbit (ISCO, e.g.~\cite{isco}) 
for the analyzed template, whichever is lower.
In our experiments, the mass ranges were varied and the number of 
templates corresponding for each mass range was calculated. 
The mass ranges and their corresponding number of templates are listed in 
Table~\ref{table:numtemplates}. In order to achieve $<5\%$ 
mismatch (i.e., $1-\rho/\rho_{\rm exp} < 5\%$ where $\rho_{\rm exp}$ 
is the expected SNR when template waveform exactly matches the signal in the data), 
thousands of templates are required~\cite{matchedfiltering} to analyze 
a data segment for mass ranges of 1.4--11 solar masses for each individual 
member of the binary. 
In the currently running search pipeline described in~\cite{ligoanalysis}, 
each data segment is made up of 256 seconds of detector data down-sampled to 
4096 Hz giving $2^{20}$ data points.
This means that thousands of FFTs, each of approximately 1 million data points, 
are required to filter one data segment through the template bank. 

\begin{table}[ht]
\caption{Mass ranges and their corresponding number of templates obtained from running the actual template generating program in the LIGO searching pipeline valid for LIGO Hanford detector H1.}
\centering
\begin{tabular}{c c}
\hline\hline
Mass Range (solar masses) & Number of templates \\
10.0 - 11 & 7 \\
9.0 - 11 & 15 \\
8.0 - 11 & 26 \\
7.0 - 11 & 48 \\
6.0 - 11 & 85 \\
5.0 - 11 & 163 \\
4.0 - 11 & 317 \\
3.0 - 11 & 718 \\
2.0 - 11 & 2111 \\
1.6 - 11 & 3734 \\
1.4 - 11 & 5222 \\
\hline
\end{tabular}
\label{table:numtemplates}
\end{table}

\subsection{The $\chi^2$ Consistency Test}
\label{sec_chi}
In order to verify the signals and reject non-Gaussian transient noise, the $\chi^2 $ consistency test~\cite{alan} is used as a time-frequency veto. 
The integral in Eq.~(\ref{eqn1}) is split into $p$ frequency bands such that 
each contributes an equal amount to the SNR, and this yeilds $p$ 
time series, $z_l(t)$, where $l$ ranges from $1$ to $p$.  In stationary 
Gaussian noise with or without a gravitational wave signal, 
the statistic~\cite{ligoanalysis}
\begin{equation}
\label{chisqeqn}
\chi^2(t) = \frac{p}{\sigma^2} \displaystyle\sum_{l=1}^p |z_l(t) - z(t) / p|^2.
\end{equation}
is a $\chi^2$-distributed random variable with $\nu = 2p-2$ degrees of
freedom.  Transient departures from Gaussian noise that are poor matches
for gravitational wave templates, or ``glitches'', are associated with
large values of the $\chi^2$ statistic, and this can be used to reject such
noise events~\cite{ligoanalysis}.

\subsection{Motivation of GPU-acceleration}
As the above discussion implies, much computing power is required to search for GWs from these sources. 
With current technology, more than 50 CPU cores are required to finish the detection phase of the analysis within the duration of the data.
The $\chi^2$ waveform consistency test requires another 50 CPU cores processing power in order to complete the analysis in real-time.
Furthermore, determination of sky directions of these sources requires hours to days of CPU-core time.  The long time scales required for detection, verification, and localization pose a serious problem for prompt follow-up observations of these sources by optical telescopes. 
Such follow-up observations are expected not only to provide firm proof that a gravitational wave has been detected but also to provide insight into the physics associated with the events.  Much faster processing is therefore required for real-time detection and determination of source directions, 

In this paper, we propose a cost-effective and user-friendly alternative to reduce the computational cost in GW searches by using the graphics processing unit (GPU) to accelerate the data processing. 
The on-going searches for gravitational-wave signals from detector data are ideal for these massively parallel processors. 
This is due to the fact that the same algorithm is applied to different data segments independently and repeatedly and that the latency of transferring data between GPU and CPU is negligible compared to analysis time. 
We report here the result of a first test, using a GPU in a modified existing data analysis pipeline described in~\cite{ligoanalysis} to search for GW signals from coalescing binaries of compact objects (denoted the {\it inspiral search pipeline}). A previous report can be found in \cite{shinkeethesis}.

\section{Graphics Processing Units and CUDA}
\label{gpu}
Graphics processing units (GPUs) were originally designed to render detailed real-time visual effects in computer games.  
The demands for GPUs in the gaming industry have enabled GPUs to become low-cost but very efficient computing devices.  
Due to the nature of its hardware architecture, it is advantageous to use GPUs 
for solving parallel problems that fit the single-instruction-multiple-data 
(SIMD) model.  While the capability of GPUs in high performance computing has been 
recognized since 2005~\cite{nbody1}, general purpose GPU (GPGPU) parallel computing 
has really become viable only recently. This is due to the release of the C-programming 
interface CUDA (Compute Unified Device Architecture) by NVIDIA Corporationr
in February 2007 \cite{cudaprog}.
The introduction of CUDA enabled scientists in a much broader community to program on GPUs by calling C-libraries.  Previously, one would have to translate a general problem into graphical pixel models in order to make use of the GPUs.  Remarkable speed-ups of up to a factor of hundreds have been reported in many applications including those of important astronomical applications~\cite{nbody1,harrisska}. A sizable CUDA library is now available for basic scientific computations.  
This includes linear algebraic computation, FFT and tools for Monte-Carlo simulation. The use of GPGPU techniques as an alternative to distributed computer clusters has also become a real possibility.

One successful application of GPU acceleration was in the computation of molecular dynamics. Anderson, Lorenz and Travesset~\cite{mdgpu} implemented CUDA algorithms to handle the core calculations for molecular dynamics.   
Anderson et al. in fact slightly altered one of the core algorithms of molecular dynamics for CUDA so that  some of the calculations will be repeated instead of accessing the same information from memory.
This avoided the problem of inefficiency of CUDA in accessing random memory locations. The CUDA program running in a system with one single GPU and one single CPU was found to be performing at equivalent level of a fast computer cluster with 36 cores. Such a cluster 
consumes 
more power compared to a single computer with a GPU~\cite{mdgpu}.

There exist several studies of CUDA implementations in the field of astronomy. Belleman et al.~\cite{nbody2} studied the CUDA implementations of N-body simulations, following the studies of Zwart et al.~\cite{nbody1} who used GPUs in N-body simulations before the release of CUDA.  
The CUDA implementation of N-body simulations developed by Belleman et al. was able to achieve up to 100 times speed-up compared to that of CPU. Harris et al.~\cite{harrisska} conducted the CUDA implementations for calculating the signal convolutions for a radio astronomy array.  
In this application, signals from each antenna are combined using convolutions.  This allows an array of antennas be used to achieve high angular resolution. The CUDA implementation of this process showed two orders of magnitude speed-up.  The use of GPUs in GW data analysis has not been reported before this writing (see~\cite{gwgpu} for an earlier proposal). 

\subsection{Crucial elements regarding GPU-acceleration}
\label{gpu_explained}
Programming for the GPU with CUDA is different from general purpose programming on the CPU due to the extremely multi-threaded nature of the device. For an algorithm to execute efficiently on the GPU, it must be cast into a data-parallel form with many thousands of independent threads of execution running the same lines of code, but on different data.   
Because of this simultaneous execution, one thread cannot depend upon the output of another, which can pose a serious challenge when trying to cast some algorithms into a data-parallel implementation. 

Specifically, in CUDA, these independent threads are organized into blocks which can contain anywhere from 32 to 512 threads each, but all blocks must have the same number of threads. Each block executes identical lines of code and is given an index to identify which piece of the data it is to operate on.  
Within each block, threads are numbered to identify the location of the thread within the block.  Any real hardware can only have a finite number of processing elements that operate in parallel. In particular, a single GeForce 8800 Ultra GPU that was used in our work contains 16 multiprocessors. 
A single multiprocessor can execute a number of blocks concurrently (up to resource limits) in warps of 32 threads.

CUDA programs are divided into 2 parts, host functions and kernel functions. The host functions are the code that run in the CPU and are able to invoke kernel functions. 
Kernel functions are the code that run in the GPU. These functions are automatically executed by the threads in each block of the GPU. 
CUDA programmers need to specify the number of blocks and the number of threads per block for the kernel functions. Threads in the same block can be synchronized (hold and wait until all threads have finished previous tasks) and communicate (access the data or output of other threads). 
However threads in different blocks cannot be synchronized. This imposes a great restriction in CUDA programming. We need to carefully choose the number of blocks and threads to obtain the highest performance.


The memory structure of a GPU is organized in a convenient way. There is a global memory (768 MB for GeForce 8800 Ultra) that can implement read-and-write operations simultaneously.  This hides the latency of data accessing between processors and memory. 
Meanwhile, each block of threads executing on a multiprocessor has access to a smaller but faster shared memory (16 KB for GeForce 8800 Ultra GPU).  Proper use of the threads and memory access are therefore crucial for optimization of the performance of GPU-computing.  It is also necessary to copy data between CPU and GPU. 
This introduces a latency that needs to be taken into account when optimizing the GPU-computing. Specifically, this means that the algorithms must be designed so that the computational time is much larger than the latency. 


\section{Application of GPUs to Inspiral Search Pipeline}
\label{tech}
The search pipelines for GWs from coalescing compact binaries have been developed since 2000~\cite{firstinspiral}. The current pipeline~\cite{ligoanalysis} has been used for the past five successful science runs on real data from the LIGO detectors~\cite{searchinspirals} and the source code is publically available in the LSC Algorithm Library (LAL)~\cite{laldaswg}. 


According to our experiments, the most time-consuming part
  (80\% of the total run time) of the existing search pipeline is the forward Fourier transform and its inverse.  Our implementation replaces the Fastest Fourier Transform in the West (FFTW)~\cite{fftw} used by the existing pipeline with the CUDA Fast Fourier Transform~\cite{cudaprog}, denoted as CUDA FFT in this paper.  
The CUDA FFT also provides functions that can calculate several FFTs in parallel, as in FFTW.  We identified modules in the pipeline that perform FFTs in sequence and rewrote them using batched CUDA FFTs. 

Our second task was to accelerate the $\chi^2$ waveform consistency test described 
in section~\ref{sec_chi}. This is by far the most computational-intensive module in the pipeline. 
Within the program, the most time-consuming part lies in a loop of FFTs that 
operate on different data segments in series, and a double loop that calculate 
the $\chi^2$ statistics from the output of the FFTs.  
We took advantage of data parallelism by copying large segments of data into the 
global memory of the GPU card, and perform the $\chi^2$ calculation in parallel 
on these data segments.

In summary, our applications of the GPUs on the inspiral 
search pipeline includes the use of the existing CUDA FFTs 
for SNR and $\chi^2$ calculations, and the direct 
implementation of parallel computation for the $\chi^2$ calculation.  
A stand-alone version of the inspiral search pipeline 
(that normally runs on computer clusters) was run on a single core 
of a Dell Inspiron 530 computer with a 2.5 GHz Intel Core 2 Quad 9300 CPU.  
The GPU card used for timing comparison is an NVIDIA GeForce 8800 Ultra 
installed in the same computer, using CUDA version 1.1. 
Stationary coloured Gaussian noise with a spectrum matching 
the Initial LIGO Science Requirement Document~\cite{ligosrd} for Hanford detector H1
was used for the test.
In all our tests and implementation, we purposely kept 
all relevant search parameters close to a real search 
as implemented in the past few science runs. 
Further acceleration could be expected once 
we consider flexible search parameters or rewrite 
a much larger fraction of the code. 


\subsection{Implementation of the CUDA Fourier-Transform}
\label{explainfftw}
As described in the previous subsection, the GW search pipeline spends the 
majority of the time in performing FFTs. LAL uses 
Fastest Fourier Transform in the West (FFTW) for performing FFTs.
FFTW was developed by Frigo and Johnson for improving the performance of 
FFTs calculations by CPUs~\cite{fftw}. 
The main feature of FFTW is that it uses a $planner$ to learn the fastest way 
to perform FFT in a computer. This $planner$ constructs plans for executing 
FFTs in a fast way, and the plans are re-used for each execution of the FFT in 
a particular computer~\cite{fftw}. Similarly, CUDA has a complete FFT library 
developed by NVIDIA which also uses a $planner$ following the design of FFTW~\cite{cudaprog}.

Our first experiment was to compare the performance of the CUDA FFT to FFTW. 
To do this, a program that generates data sets and performs CUDA FFTs on the 
data was developed. 
The transferring of data from the host CPU to GPU (and vice versa) and the 
CUDA FFT calculation was put into a loop so that appropriate repetitions of the 
calculations could be timed. This means that the time measured is the data transfer 
time plus the CUDA FFT execution time. 
A similar program that uses FFTW instead of the CUDA FFT on the same set of data was also developed and timed. The program that uses FFTW does not need to transfer data from elsewhere.

\begin{figure}
\centerline{\includegraphics[keepaspectratio=true,height=3.4in,angle=0]{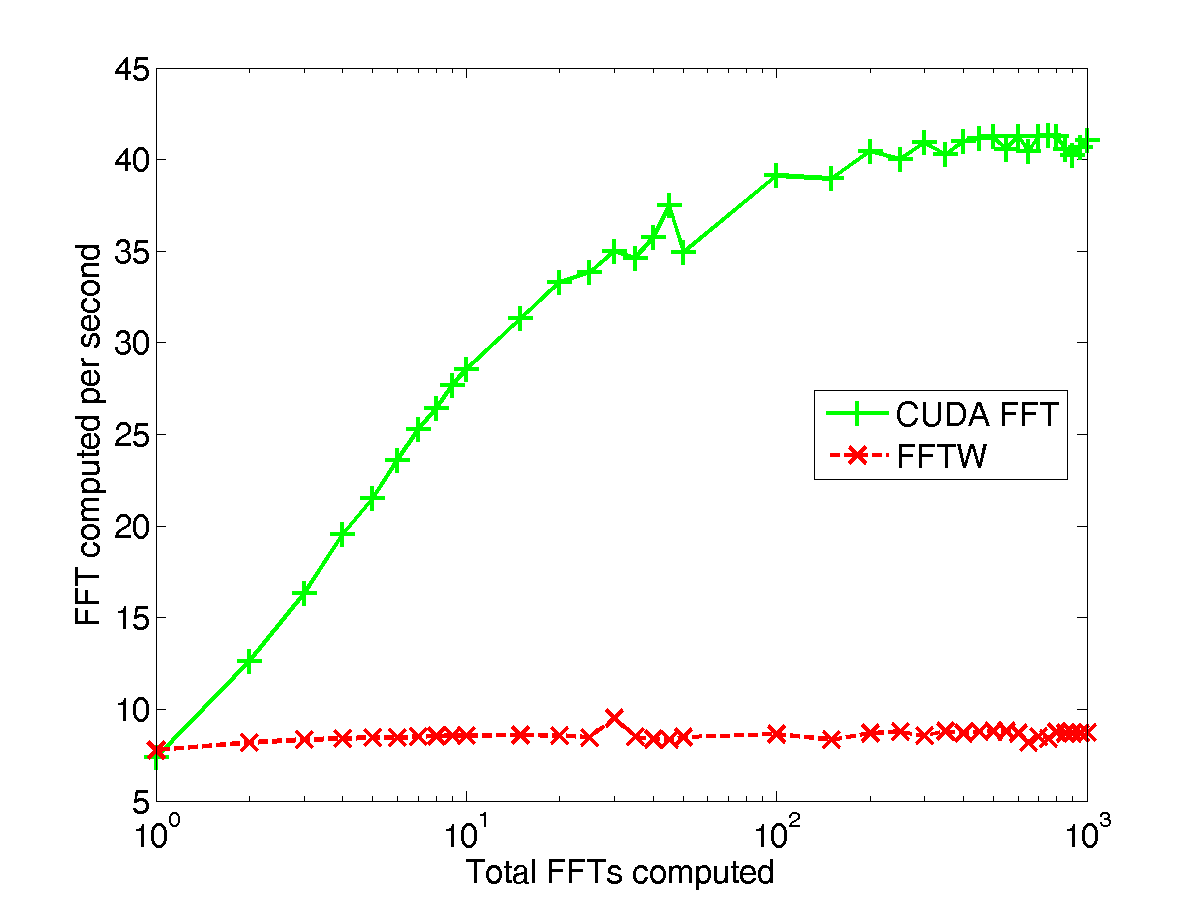}}
\caption{FFTs executed per second as a function of the total number of FFTs executed with $2^{20}$ data points each. The green solid line shows the number of FFTs executed by CUDA, while the red dashed line shows the number of FFTs executed by FFTW.}
\label{fig:1mfftpersec}
\end{figure}

\begin{figure}
\centerline{\includegraphics[keepaspectratio=true,height=3.4in,angle=0]{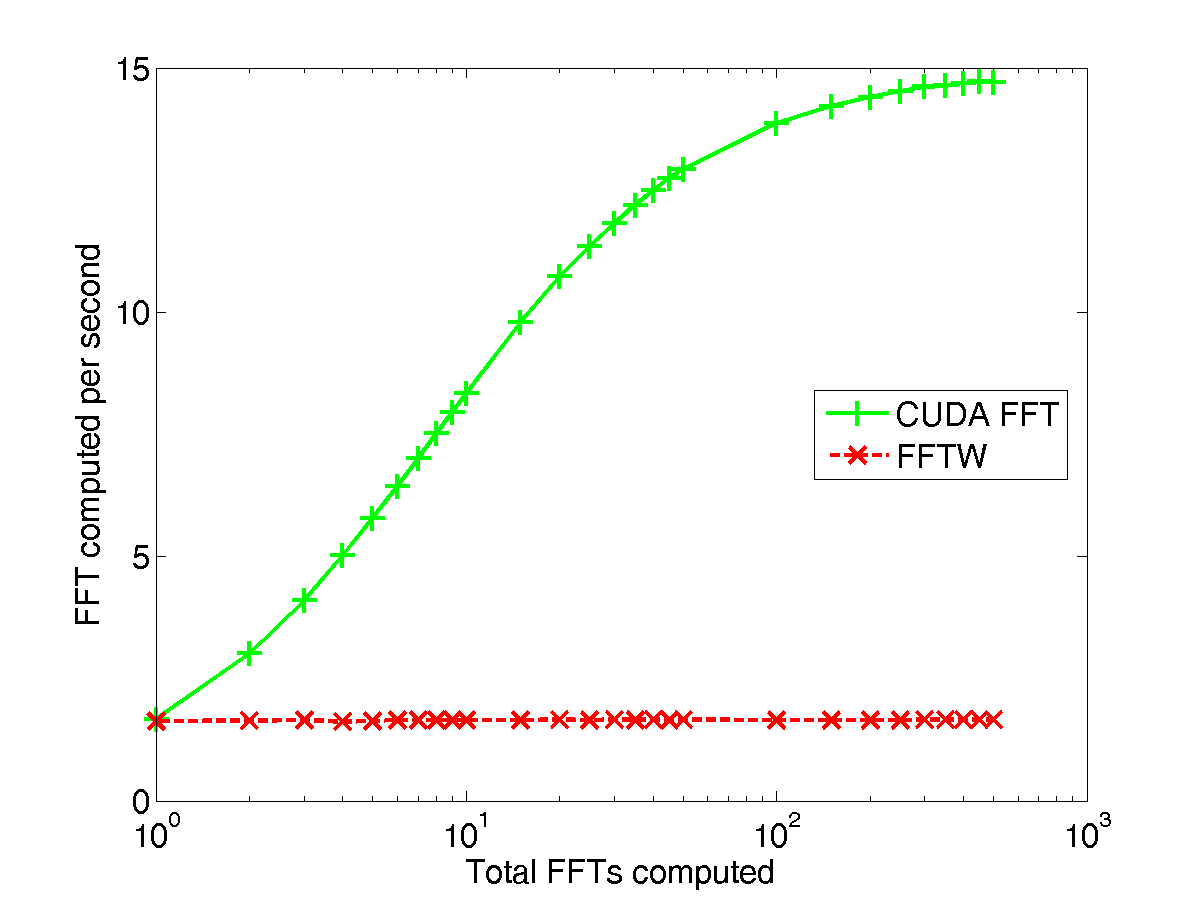}}
\caption{FFTs executed per second as a function of the total number of FFT executed with $4 \times 2^{20}$ data points each. The green solid line shows the number of FFTs executed by CUDA, while the red dashed line shows the number of FFTs executed by FFTW.}
\label{fig:4mfftpersec}
\end{figure}

The comparison of CUDA FFT and FFTW is shown in Figure~\ref{fig:1mfftpersec} 
and Figure~\ref{fig:4mfftpersec}. The graphs show the number of FFTs executed 
per second (vertical axis) against the number of FFTs executed (horizontal axis). 
The vertical axis therefore indicates the capability of the hardware. 
Higher values on the vertical axis indicate better capability of the computer in 
performing FFT operations. The performance of the CUDA FFT shows a fast rise at a 
low number of FFTs and reaches a plateau at larger numbers.  
The number of FFTs  was incremented by one for each step at the beginning to 
show clearly the initial quick rise in performance.  As the graph flattens at 
higher number of FFTs,  the increment is fifty for each step. In the experiment, 
the time for transferring data from the host CPU memory to GPU memory was found 
to be negligible compared to the CUDA FFT execution time.


An interesting feature shown in  Figure~\ref{fig:1mfftpersec} and 
Figure~\ref{fig:4mfftpersec} is the poor performance of the CUDA FFT 
for small numbers of FFTs. As the experiments were carried out, 
we found that there is an initialization period of ``warming up'' time 
associated with the GPU before any executions could be performed 
on it (even before we can start transferring data to the GPU). 
Our GPU was tested in CUDA version 1.1 with a program that 
repeatedly allocates memory in the GPU.
We found that there is generally a huge delay in allocating the first memory
space, in the order of 100 ms.
The next memory allocation takes less than 1 ms.
Therefore, the GPU does not perform well when the number of FFTs 
performed is small where the initialization period is a significant 
fraction of the total time.  
In fact, if we perform only one FFT, then the CUDA FFT is slower than FFTW.  
However, there is a quick rise in the performance of the CUDA FFT at the beginning 
when the number of FFTs is increased. Both the CUDA FFT performance curves 
flatten out at about 1000 total FFTs when the ``warming up'' time is much 
smaller than the total execution time. 

Figure~\ref{fig:1mfftpersec} shows that CUDA FFT can execute about 40 FFTs
 per second for $2^{20}$ data points, while FFTW can execute about 8 FFTs 
per second. That means CUDA FFT is performing 5 times faster than FFTW.  
 Figure~\ref{fig:4mfftpersec} shows about 7.5 times speed-up for 
$4 \times 2^{20}$ data points.   In the inspiral search pipeline, 
the FFTs are performed on $2^{20}$ data points.   Our results indicate 
that more speed-up can be achieved if more data points are used. 
 Overall, it is only worth the effort to use CUDA FFTs if a large number 
of FFTs are to be executed.

We developed an interface of the CUDA FFT to be used by the GW search community 
in general. This was done by adding the interface into LAL for calling the CUDA FFT library 
which can be manually activated or deactivated by LAL users. This is the first time that a GPU interface was successfully developed for 
LAL and tested with real applications.

\subsection{Acceleration using Data Parallelism of GPUs} 
We applied the GPU data parallelism to the most computationally intensive and 
time consuming stage of the gravitational-wave search pipeline, the $\chi^2$ test. 
The $\chi^2$ test splits the inspiral template into 16 pieces in the 
frequency domain, convolves the Fourier transform of the input data with 
the split
16 time series representing the contribution to the template's net SNR 
(as a function of time) from each of its 16 pieces. 
Suitably normalized, the sum of the square magnitudes of these 16 time series 
is $\chi^2$-distributed when the input data contain stationary Gaussian noise and 
a possibly-absent gravitational-wave signal, and testing for this forms the basis 
of a waveform consistency test.

\begin{figure}
\centerline{\includegraphics[keepaspectratio=true,height=3.0in,angle=0]{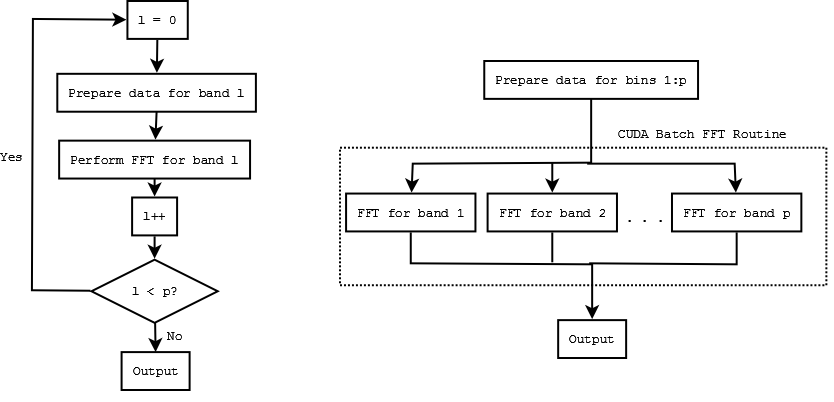}}
\caption{The comparison flow chart of the sequential FFTs and batched FFTs execution in the $\chi^2$ calculation.}
\label{fig:flowchart}
\end{figure}

The conversion of the $\chi^{2}$ test to a GPU implementation was done in two parts. 
Firstly, 16 sequential inverse FFTs were replaced with 16 parallel inverse FFTs. 
This part was implemented by calling existing CUDA functions from the host code.  The comparison of this implementation to the original one is shown in Figure~\ref{fig:flowchart}.  
Secondly, we implemented the GPU data parallelism on the $\chi^2$ test described 
in section~\ref{sec_chi}. 
Table~\ref{codetable} shows the $\chi^2$ implementation in C and the GPU-accelerated 
implementation. $a_{i,l}$ and $b_{i,l}$ represent the real and imaginary part of $z_l(t) - z(t)/p$ 
in Eq.~(\ref{chisqeqn}) respectively. 
$N$ is the number of data points being analyzed, $i$ ranges from $0$ to $N-1$ and 
$p$ is the total number of frequency bands. 
In the original implementation, a double loop was used to 
calculate $\chi^2$ values sequentially with $p=16$ and $N=2^{20}$. A total of $16 \times 2^{20}$ $\chi^2$ 
values were thus calculated independently.  For each time $t$, values from all 16 frequency bands were then added (c.f. Eq.~(\ref{chisqeqn})), yielding a total of $2^{20}$ outputs.  
In the CUDA implementation,  we replaced this double loop with a single loop in 
the parallel threads. 
This part was implemented 
in a custom kernel function where $4 \times 2^{10}$ blocks of $2^{9}$ threads each 
were used.  Each thread calculates eight $\chi^2$ values and sums them together sequentially.  
Adjacent threads were used to calculate $\chi^2$ values of the same frequency band.  
The results from every two adjacent threads are then summed at the end of this single loop 
after a synchronization was executed.  This approach was found to give optimal performance. 
The thread numbers are chosen to be multiples 
of the GPU warp size 32 (explained in 
Section~\ref{gpu_explained}) and able to divide the loop number exactly.

\begin{table}[ht]
\caption{Algorithm comparison of $\chi^2$ implementation in C and CUDA .}
\centering
\begin{tabular}{|l|l|}
\hline
Original C implementation & CUDA implementation \\ [1ex]
\hline
\verb!for(i=0; i<N; i++){!&Thread $i_1$: \\
\verb!	for(l=0; l<p; l++){!&\verb! for(l=0; l<p/2; l++)! \\
\verb!	  ! $\chi^2_i$ \verb! += !$a^2_{i,l} + b^2_{i,l}$&\verb!    !$\chi^2_{i_1}$ \verb!+=! $a^2_{i,l} + b^2_{i,l}$;\\
\verb!	}! & \\
\verb!}! & Thread $i_2$: \\
&\verb! for(l=p/2; l<p; l++)! \\
&\verb!    !$\chi^2_{i_2}$ \verb!+=! $a^2_{i,l} + b^2_{i,l}$;\\
& \\
& Synchronizing all threads:\\
& \verb!  !$\chi^2_i$ \verb!+=! $\chi^2_{i_1} + \chi^2_{i_2}$;\\ [1ex]
\hline
\end{tabular}

\label{codetable}
\end{table}


\subsection{Results}
\label{results}
\begin{figure}
\centerline{\includegraphics[keepaspectratio=true,height=3.4in,angle=0]{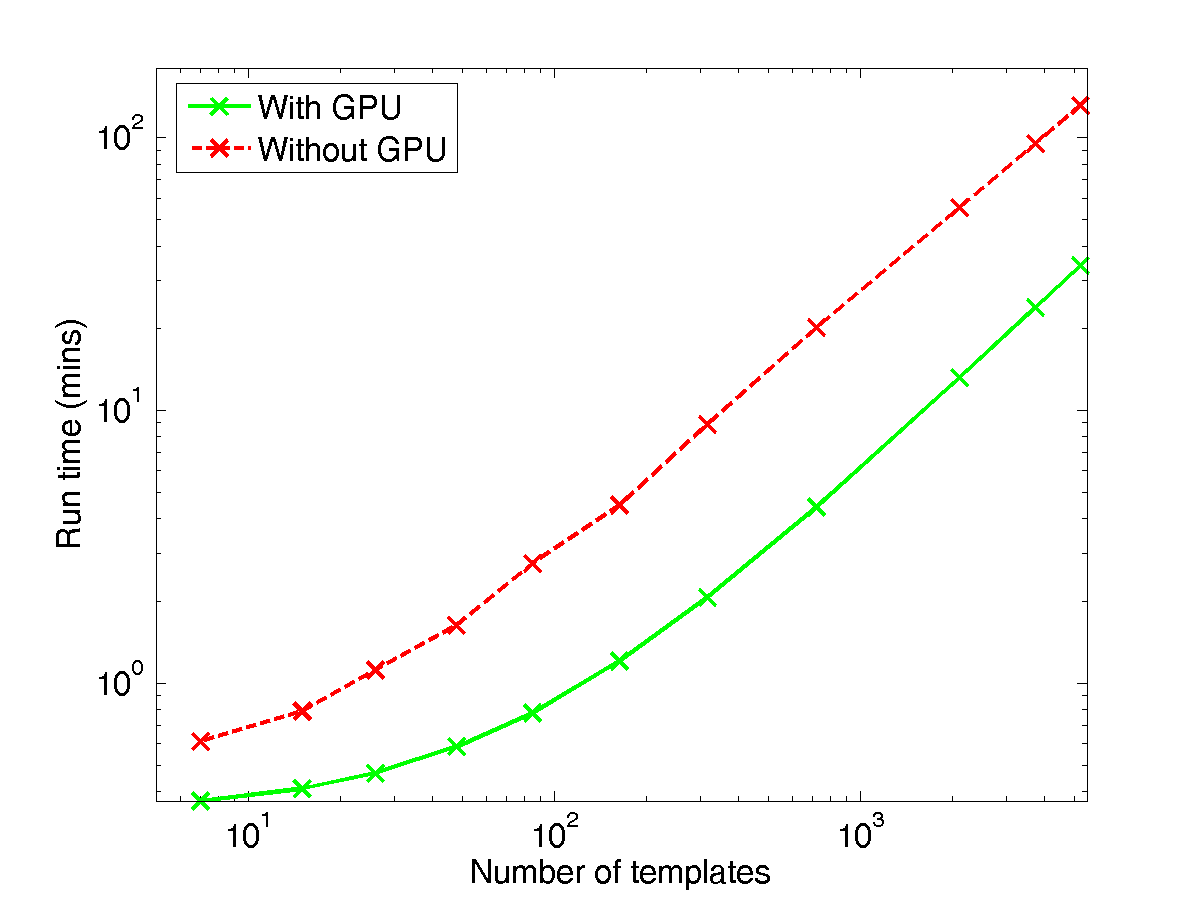}}
\caption{The run time of the inspiral searching pipeline without performing the $\chi^2$ test. The green solid line shows the run time of inspiral search with GPU, while the red dashed line shows the run time of inspiral search without GPU.}
\label{inspiraltimes}
\end{figure}

\begin{figure}
\centerline{\includegraphics[keepaspectratio=true,height=3.4in,angle=0]{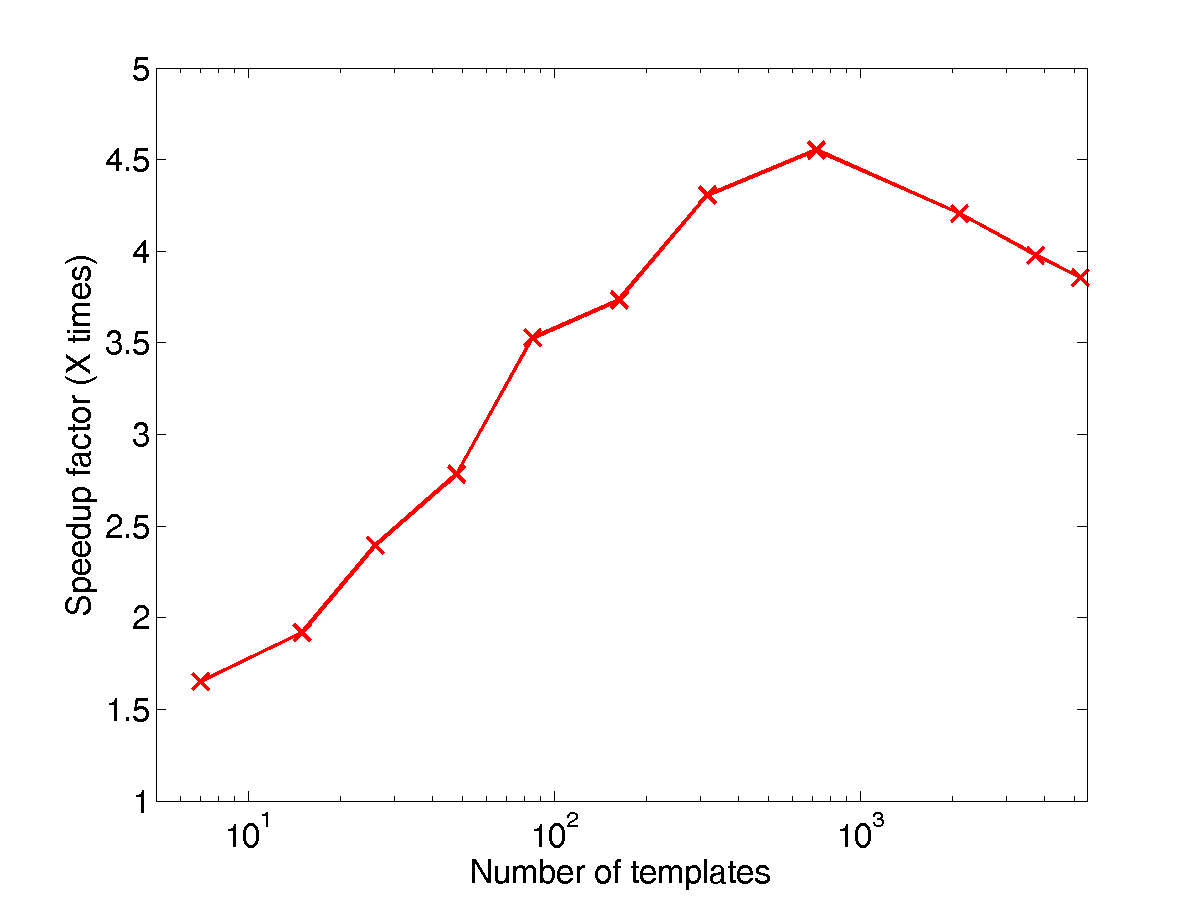}}
\caption{The speed-up factor, calculated as run time without GPU divided by run time with GPU.}
\label{inspiralfactor}
\end{figure}

\begin{figure}
\centerline{\includegraphics[keepaspectratio=true,height=3.4in,angle=0]{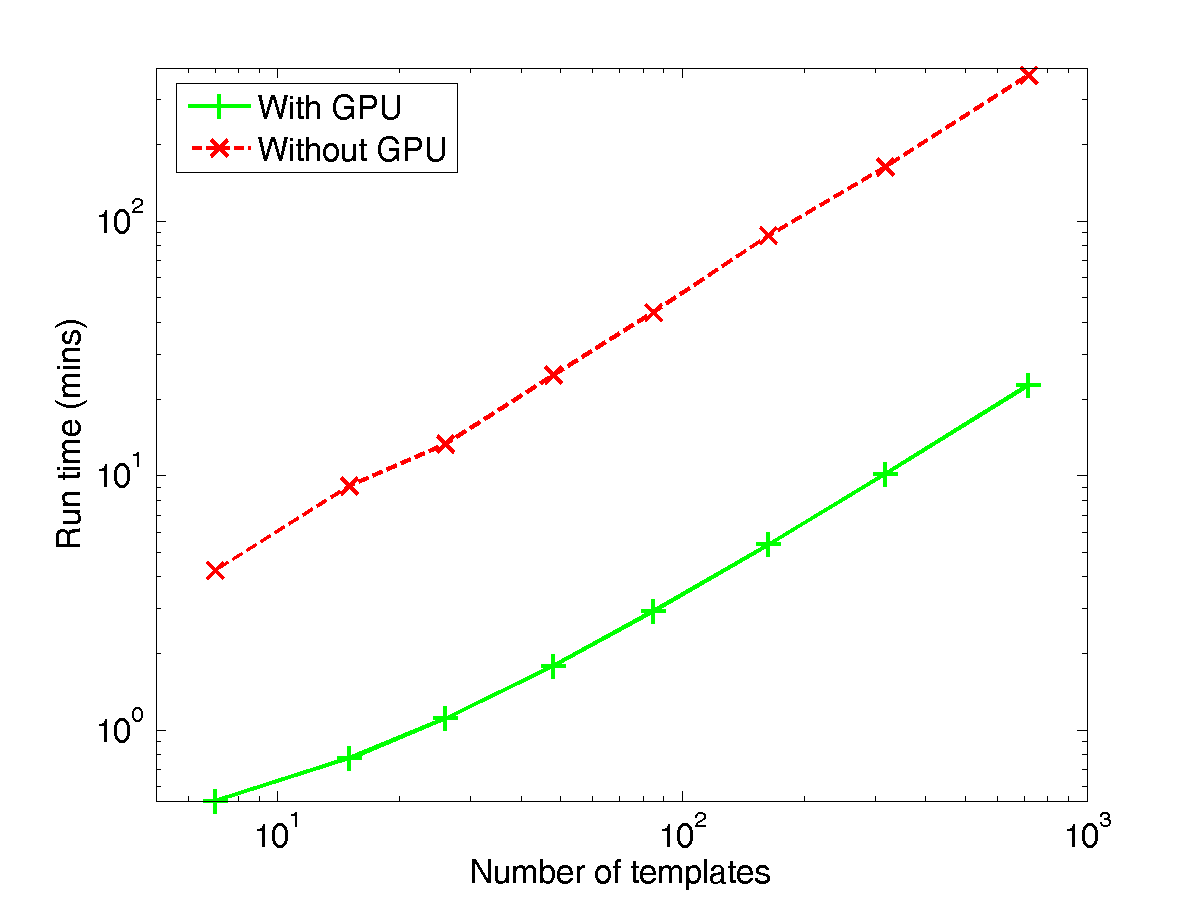}}
\caption{The run time for executing the inspiral search with $\chi^2$ veto enabled, both with and without GPU acceleration. The green solid line shows the run time of inspiral search with the GPU, while the red dashed line shows the run time of inspiral search without the GPU. }
\label{fig:chisqtime}
\end{figure}

\begin{figure}
\centerline{\includegraphics[keepaspectratio=true,height=3.4in,angle=0]{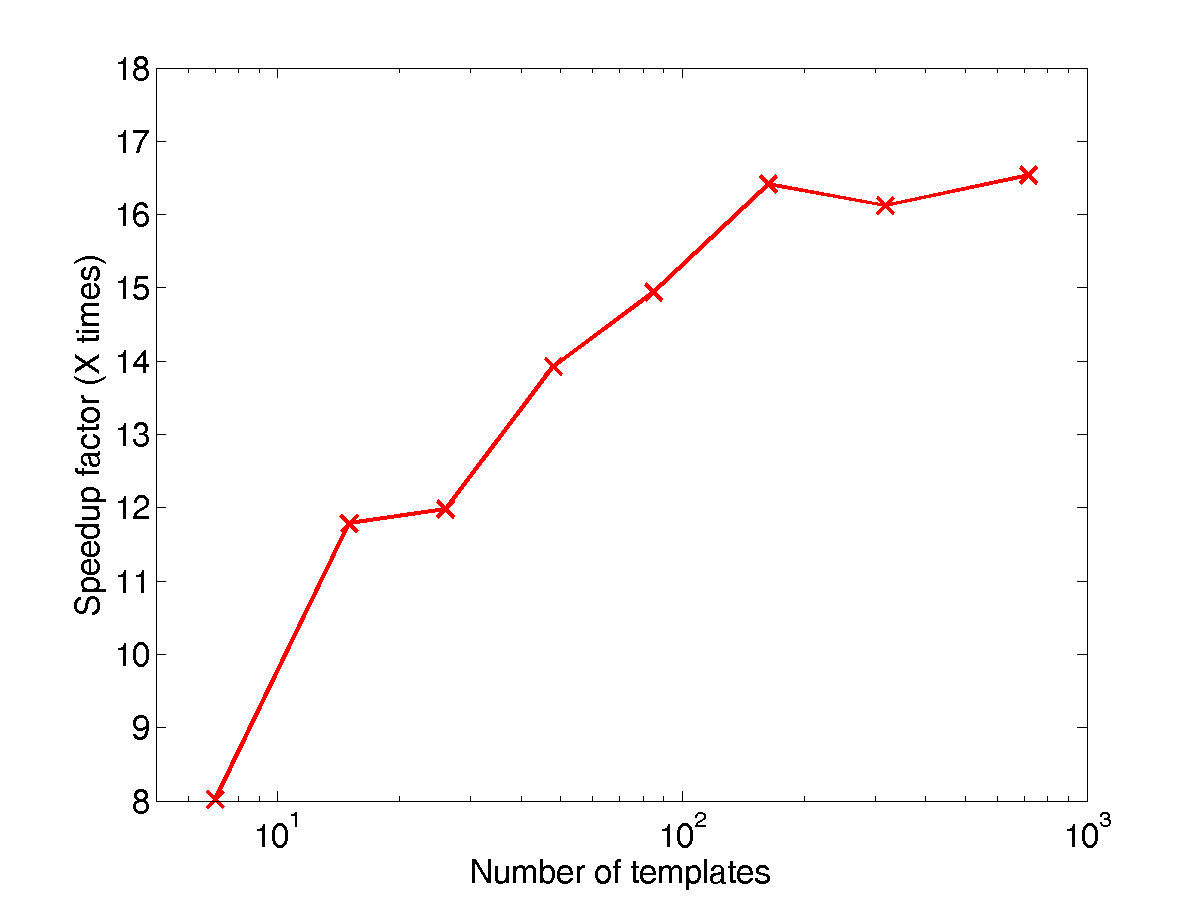}}
\caption{The speed-up factor (see text in Section~\ref{results}) using the CUDA implementation of the $\chi^2$ calculation.}
\label{fig:chisqspeedupfac}
\end{figure}

The timing results of our implementation of CUDA FFT and data parallelism for $\chi^2$ implementation in the inspiral search pipeline are shown in Figure~\ref{inspiraltimes}, Figure~\ref{inspiralfactor}, Figure~\ref{fig:chisqtime} and Figure~\ref{fig:chisqspeedupfac}.  
About 4x speed-up 
can be achieved by simply enabling this CUDA FFT interface for LAL (Figure~\ref{inspiraltimes} and Figure~\ref{inspiralfactor}). 
The run time of 
the inspiral pipeline was shown in Figure~\ref{inspiraltimes} and the speed-up factor
was shown in Figure~\ref{inspiralfactor}.

In Figure~\ref{fig:chisqtime}, the vertical axis shows the run time of the inspiral search pipeline while the horizontal axis shows the number of templates used for the search. It is shown that, at about 700 templates, the inspiral search using the original $\chi^2$ implementation took about 6 hours to complete, while it required only about 20 minutes to complete with our GPU implementation.
 
The speed-up factor of the GPU implementation compared to the CPU-only implementation is shown in Figure~\ref{fig:chisqspeedupfac}.  The vertical axis shows the speed-up factor --- the run time of the original CPU-only implementation divided by the run time of the GPU implementation. About 16 times speed-up was observed. 
This means that the number of computers needed to perform the analysis in the same amount of time can be significantly reduced. A normal computer with integrated graphics should consume about $\unit{220}{\watt}$ of power, or $\unit{3520}{\watt}$ for 16 single core computers, or $\unit{880}{\watt}$ for 4 quad core computers. In comparison, a single computer with GeForce 8800 Ultra consumes about $\unit{340}{\watt}$ of power. We could save some hardware costs and also reduce power consumption by a significant amount. 

An accuracy test was performed by calculating the fractional difference between 
the outputs produced by the new inspiral search with CUDA FFT including 
the parallel $\chi^2$ implementation and the original inspiral search with 
FFTW in the mass range of 3.0 - 11.0 solar masses. 
About $5 \times 10^{4}$ events were identified from the data, each with measured 
SNR and $\chi^{2}$ values. We found that more than 99\% of the SNRs had less 
than 0.03\% difference, while 99\% of the $\chi^2$ values had less than 0.5\% difference.

\section{Conclusion}
\label{concl}
We have shown that GPUs can significantly improve the speed of gravitational wave data analysis. A speed-up of 4 to 5 fold in the existing inspiral search pipeline can be achieved by simply enabling the CUDA FFT. Note that the CUDA FFT has already been introduced to LAL, meaning that other GW search pipelines can use it provided GPUs are available.
We achieved a 16 fold speed-up in total by using a specially-written parallel GPU implementation of the $\chi^{2}$ test, a waveform consistency test used within the pipeline. We expect further speed-ups if we are allowed to change some of the search parameters.  For instance, if we change the number of data points for FFTs from the currently used 1 million to 4 millions, another factor of 2 speed-up can be achieved.  Also, further acceleration is expected if we replace more components in the pipeline with specially-written GPU implementations.

Our experiments were performed using a single GPU, while current new personal computers can be equipped with more than 3 GPUs. We would expect more than 48 fold speed-up using a 3 GPUs system when running a single-threaded search pipeline. 
Furthermore, if we can use the newest GPU on the market, which has about 1 TFLOPS of computing power, and assuming that the performance of these GPUs scales linearly, we would expect more than a 100 fold speed-up in a single core desktop computer. 


\ack
We are grateful to Chad Hanna, Drew Keppel, Phil Ehrens, Stuart Anderson, Alan Weinstein, Yanbei Chen, Karen Haines, Shaun Hooper, Adam Mercer and Oliver Bock
for discussions of this work. This work is supported in part by David and Barbara Groce start-up fund at Caltech, 
the Caltech SURF program, the Alexander von Humboldt Foundation's Sofja Kovalevskaja Programme funded by the 
German Federal Ministry of Education and Research, the Australian Research Council and the WA Centres of Excellence Program.


\section{References}
\begin{thebibliography}{10}
\bibitem{pathtoenhanced} Smith, J.~R., \& for the LIGO Scientific Collaboration 2009, The path to the Enhanced and Advanced LIGO gravitational-wave detectors, Classical and Quantum Gravity, 26, 114013
\bibitem{ligoanalysis} Abbott, B., et al. 
2004, Analysis of LIGO data for gravitational waves from binary neutron stars, Physical Review D, 69, 122001 
\bibitem{coalescingbinaries} Hughes, S.~A.\ 2009, Gravitational waves from merging compact binaries, arXiv:0903.4877 
\bibitem{matchedfiltering}Owen, B.~J., \& Sathyaprakash, B.~S.\ 1999,  Matched filtering of gravitational waves from inspiraling compact binaries: Computational cost and template placement, Physical Review D, 60, 022002
\bibitem{finn} Finn, L.~S.1992, Detection, measurement, and gravitational radiation, Physical Review D, 46, 5236 
\bibitem{blanchet96} Blanchet, L.,  Iyer,B.R., Will,  C.M. ,  and Wiseman, A.G., 1996, Gravitational waveforms from inspiralling compact binaries to second-post-Newtonian order,   Classcal and Quantum Gravity, 13, 575
\bibitem{prd49} Dhurandhar, S.~V., \& Sathyaprakash, B.~S.\ 1994, Choice of filters for the detection of gravitational waves from coalescing binaries. II. Detection in colored noise, Physical Review D, 49, 1707 
\bibitem{sfformula} Damour, T., Iyer, B.~R. \& Sathyaprakash, B.~S.\ 2001, Comparison of search templates for gravitational waves from binary inspiral, Physical Review D, 63, 044023 
\bibitem{ligosrd} Lazzarini, A. \& Weiss, R.\ July 1996, LIGO Science Requirements Document (SRD), LIGO-E950018-02-E, available from: http://www.ligo.caltech.edu/docs/E/E950018-02.pdf.
\bibitem{isco}Buonanno, A.\ 2002, Gravitational waves from inspiralling binary black holes, Classical and Quantum Gravity, 19, 1267 
\bibitem{alan} Allen, B.\ 2005, $\chi^2$ time-frequency discriminator for gravitational wave detection, Physical Review D, 71, 
062001 
\bibitem{shinkeethesis} Chung, S. K., Honours thesis, School of Computer Science and Software Engineering, The University of Western Australia, 2008
\bibitem{nbody1}Portegies Zwart, S.~F., Belleman, R.~G., \& Geldof, P.~M.\ 2007, High-performance direct gravitational N-body simulations on graphics processing units, New Astronomy, 12, 641 
\bibitem{cudaprog} NVIDIA Corporation. NVIDIA CUDA Programming Guide 1.1, 2007, available from: www.nvidia.com/object/cuda\_develop.html.
\bibitem{harrisska} Harris, C., Haines, K., \& Staveley-Smith, L.\ 2008, GPU accelerated radio astronomy signal convolution, Experimental Astronomy, 22, 129 
\bibitem{mdgpu} Anderson, J.~A., Lorenz, C.~D., \& Travesset, A.\ 2008, General purpose molecular dynamics simulations fully implemented on graphics processing units, Journal of Computational Physics, 227, 5342 

\bibitem{nbody2}Belleman, R.~G., B{\'e}dorf, J., \& Portegies Zwart, S.~F.\ 2008, High performance direct gravitational N-body simulations on graphics processing units II: An implementation in CUDA, New Astronomy, 13, 103
\bibitem{gwgpu} Gy\H{o}z\H{o} Egri Feb 2008, Parallel Computation on Graphical Processor Units with an Eye on Gravitational-wave Data Analysis Needs, available from: http://www.ligo.caltech.edu/docs/T/T070308-00.pdf
\bibitem{firstinspiral} Tagoshi, H., et al.\ 2001, First search for gravitational waves from inspiraling compact binaries using TAMA300 data, Physical Review D, 63, 062001 
\bibitem{searchinspirals} Brown, D.~A., et al.\ 2004, Searching for gravitational waves from binary inspirals with LIGO, Classical and Quantum Gravity, 21, 1625 
\bibitem{laldaswg} Data Analysis Software Working Group, LAL Home Page, available from: https://www.lsc-group.phys.uwm.edu/daswg/projects/lal.html.
\bibitem{fftw} Frigo, M., and Johnson, S. G. FFTW. MIT, available from: www.fftw.org.
\end{thebibliography}
\end{document}